\documentclass[journal=jpclcd,manuscript=letter]{achemso}
\usepackage{amsmath,amssymb}
\usepackage{float}
\usepackage[version=3]{mhchem} 
\usepackage{graphicx}
\usepackage{adjustbox}
\usepackage{amsmath,amssymb}
\usepackage{mathrsfs}
\usepackage{color}
\usepackage{multirow}
\usepackage{xcolor}
\usepackage{array}
\usepackage{afterpage}
\usepackage{upgreek}
\usepackage{enumerate}
\usepackage[normal]{subfigure}
\usepackage[figurename={Figure}]{caption}
\usepackage{float}
\usepackage{gensymb} 

\usepackage{titlesec}
\titlelabel{\thetitle. }

\SectionNumbersOn

\author{Ankita Phutela}
\email{Ankita@physics.iitd.ac.in [AP]}
\author{Sajjan Shoeran}
\author{Saswata Bhattacharya}
\email{saswata@physics.iitd.ac.in [SB]}
\phone{+91-11-2659 1359}
\fax{+91-11-2658 2037}
\affiliation[Indian Institute of Technology Delhi]
{Department of Physics, Indian Institute of Technology Delhi, New Delhi, India}
\title[An \textsf{achemso} demo]
{Strain-driven topological quantum phase transition in the family of halide perovskites}
\begin{document}
\begin{center}
{\Large \bf Supplementary Information}\\ 
\end{center}
\begin{enumerate}[\bf I.]
\item Band profile of FAPbI$_3$ using PBE+SOC and G$_0$W$_0$@HSE06+SOC $\epsilon_{xc}$ functionals.
\item Structural stability of cubic and pseudocubic FAPbI$_3$.
\item Continuous TPT in cubic CsPbI$_3$.
\item TPT in cubic and pseudocubic MAPbI$_3$.
\item Crystal structures and structural stability of Cs$_{0.5}$MA$_{0.5}$PbI$_3$, Cs$_{0.75}$MA$_{0.25}$PbI$_3$ and Cs$_{0.25}$MA$_{0.75}$PbI$_3$.
\item TB parameters for Cs$_{0.5}$MA$_{0.5}$PbI$_3$. 
\end{enumerate}
\vspace*{12pt}
\clearpage
{\textbf{Band profile of FAPbI$_3$ using PBE+SOC and G$_0$W$_0$@HSE06+SOC $\epsilon_{xc}$ functionals.}}\\
\begin{figure*}[htp]
	\includegraphics[width=0.7\textwidth]{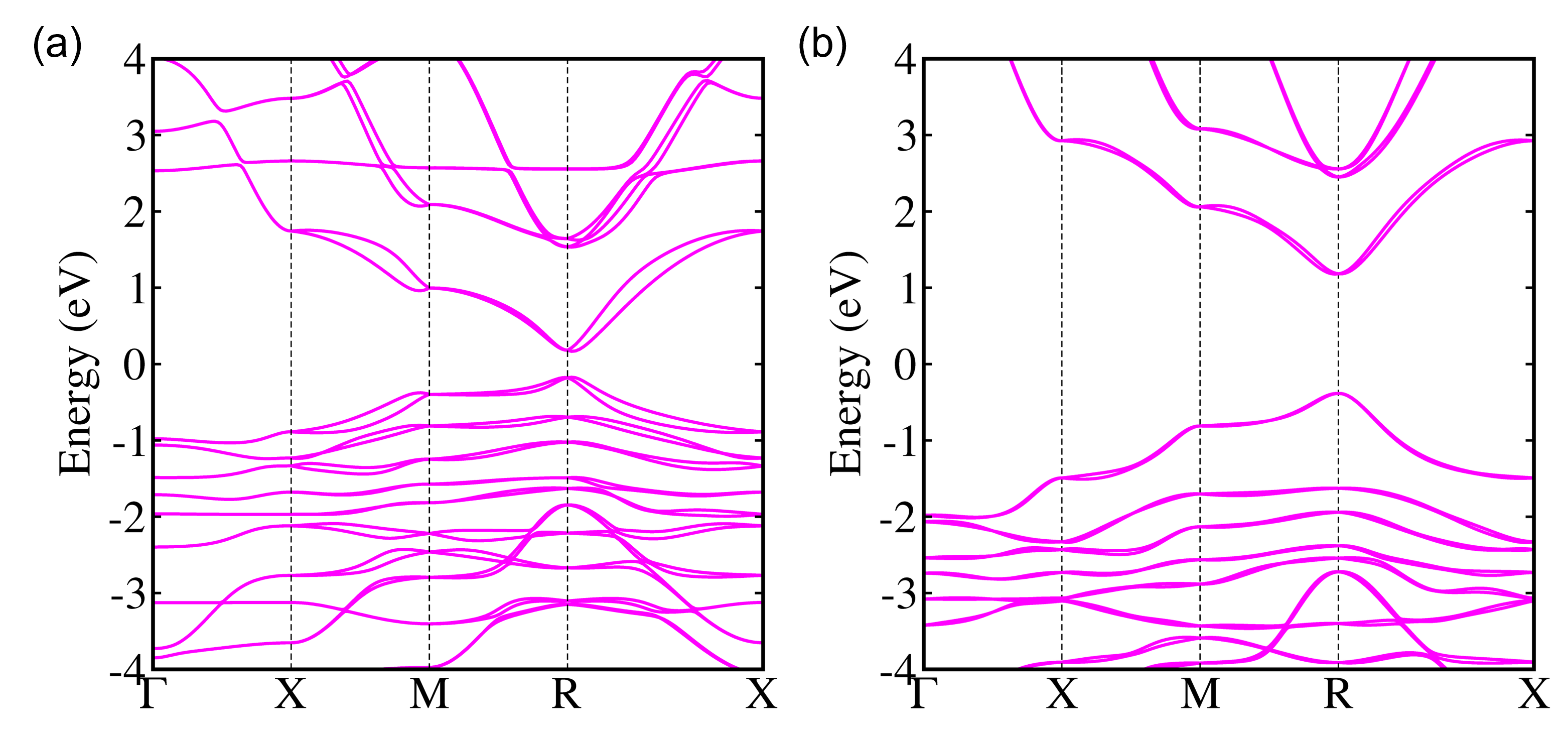}
	\caption{Band structure of FAPbI$_3$ calculated using (a) PBE+SOC and (b) G$_0$W$_0$@HSE06+SOC.}
	\label{p1}
\end{figure*}
\newpage

\begin{figure*}[h]
	\includegraphics[width=0.85\textwidth]{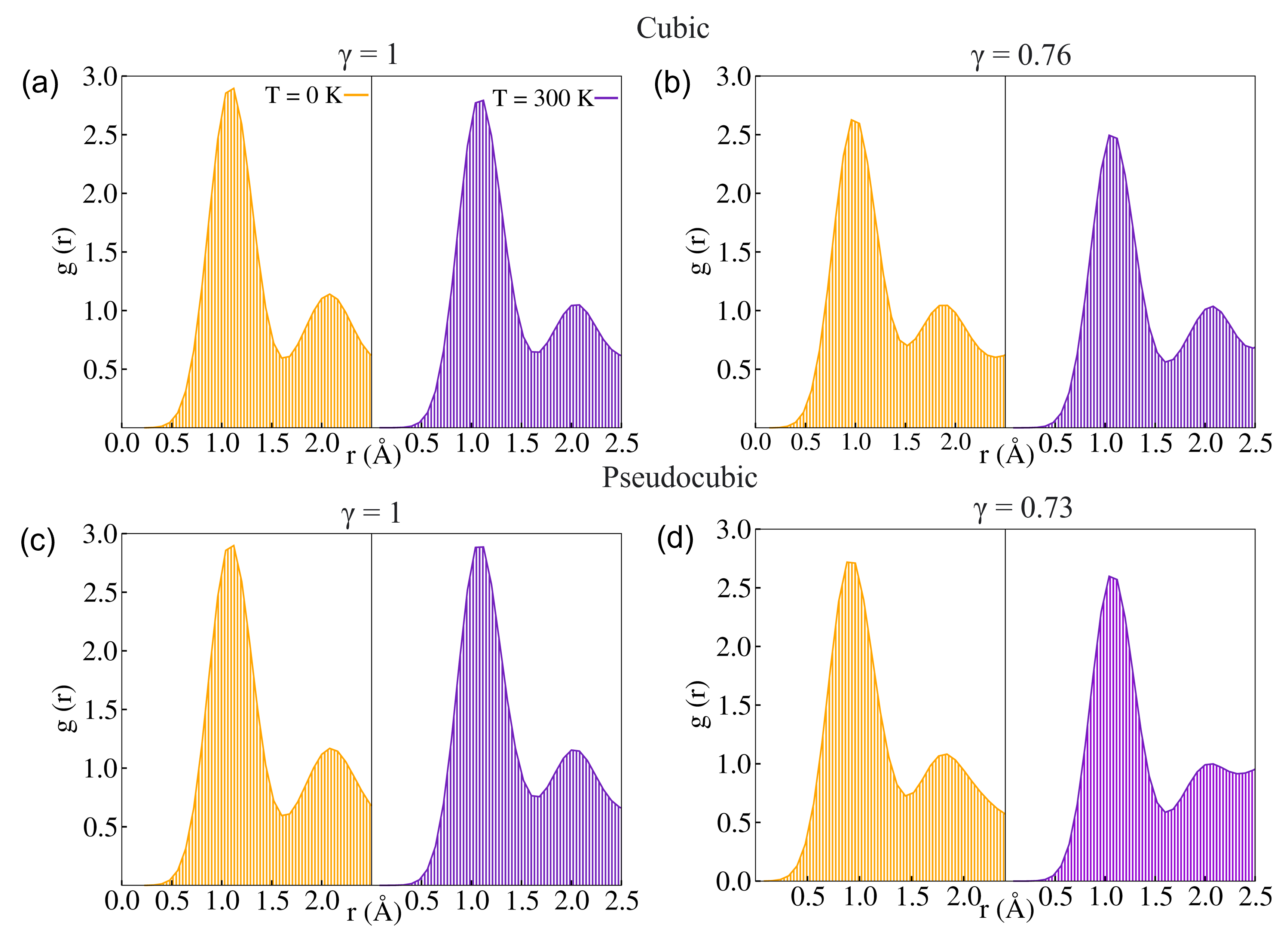}
	\centering
	\caption{Radial distribution function for different phases of FAPbI$_3$ namely (a) cubic, (b) strained cubic ($\gamma$ = 0.76), (c) pseudocubic and (d) strained pseudocubic ($\gamma$ = 0.73) at T = 0K and T = 300K.}
	\label{pic3}
\end{figure*}

{\textbf{Structural stability of cubic and pseudocubic FAPbI$_3$.}}\\
We have checked the structural stability of cubic and pseudocubic FAPbI$_3$ along with the  strained configurations using ab \textit{initio} molecular dynamics (AIMD). For this, we have obtained the radial distribution function (g(r)) at T = 0K and T = 300K. We have observed that the nature of radial distribution function remains the same at room temperature for the nearest neighbors (see Figure \ref{pic3}). Therefore, we validate that the configurations are structurally stable at 300K even after undergoing compression. This makes them useful in realizing the practical applications.

\newpage

{\textbf{Continous TPT in cubic CsPbI$_3$.}}\\
\begin{figure*}[htp]
	\includegraphics[width=0.75\textwidth]{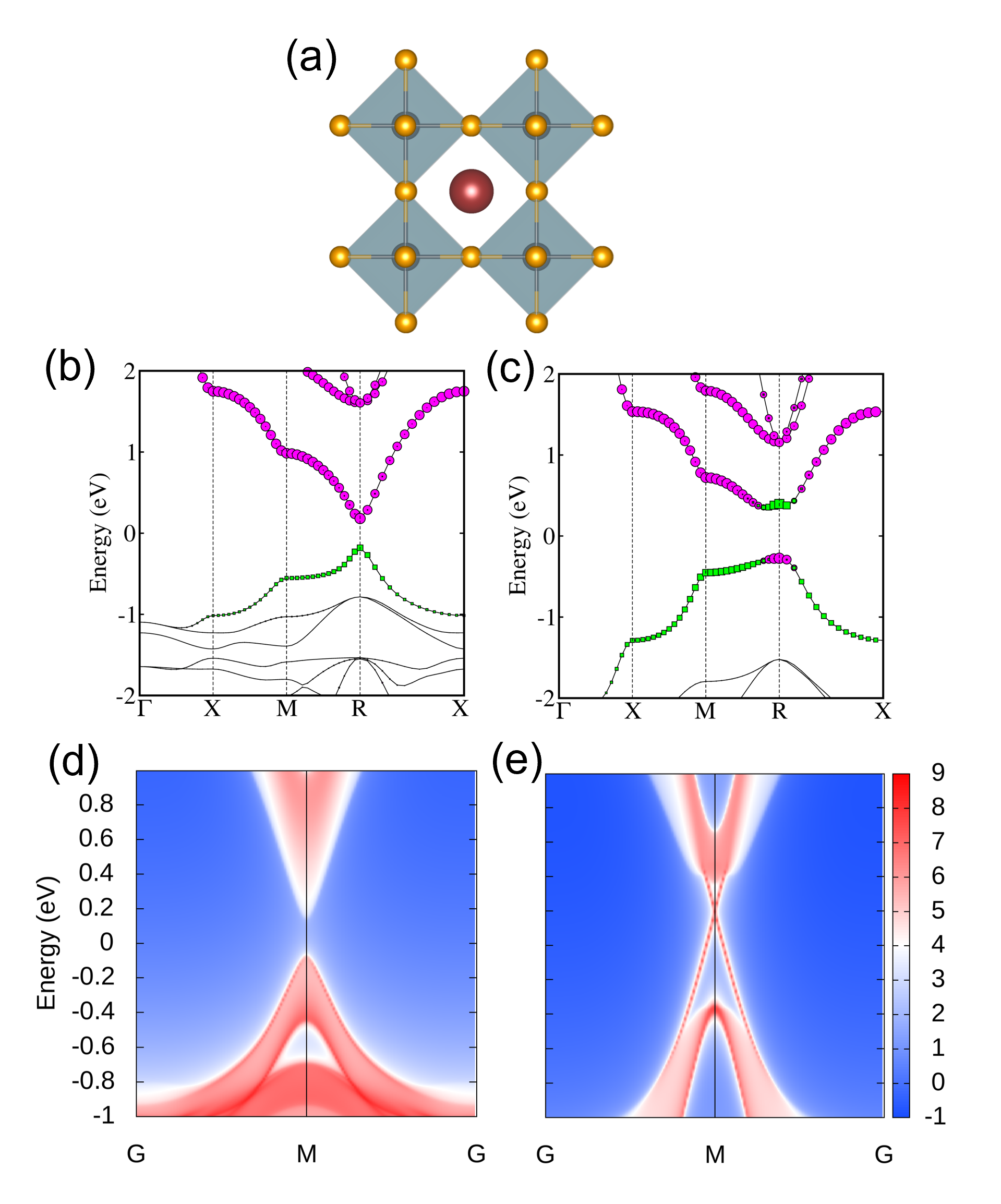}
	\caption{(a) Crystal structure of cubic CsPbI$_3$. Bulk band structure (b) without compression with Pb-\textit{p} (magenta) contribution at CBM and Pb-\textit{s} (green) contribution at VBM, and (c) with compression $\gamma$= 0.76, showing band inversion at R point. Surface band structure (d) without, and (e) with compression, showing a continuous tranistion from normal to topological state.}
	\label{1}
\end{figure*}

\newpage

{\textbf{TPT in cubic and pseudocubic MAPbI$_3$.}}\\
\begin{figure*}[htp]
	\includegraphics[width=0.7\textwidth]{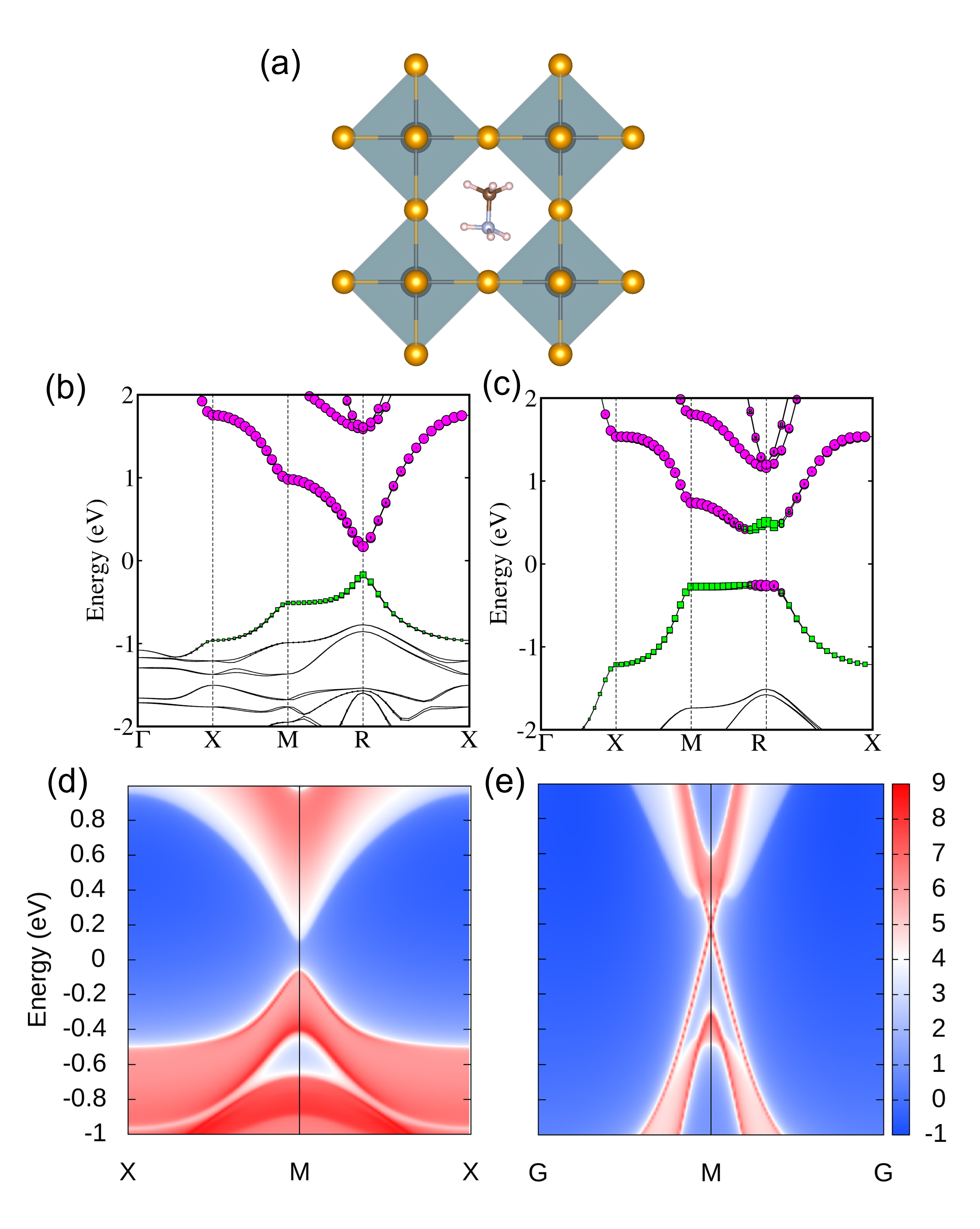}
	\caption{(a) Crystal structure of cubic MAPbI$_3$. Bulk band structure (b) without compression with Pb-\textit{p} (magenta) contribution at CBM and Pb-\textit{s} (green) contribution at VBM, and (c) with compression $\gamma$= 0.73, showing band inversion at R point. Surface band structure (d) without, and (e) with compression $\gamma$= 0.73, showing a continuous tranistion from normal to topological state.}
	\label{2}
\end{figure*}


\begin{figure*}[htp]
	\includegraphics[width=0.7\textwidth]{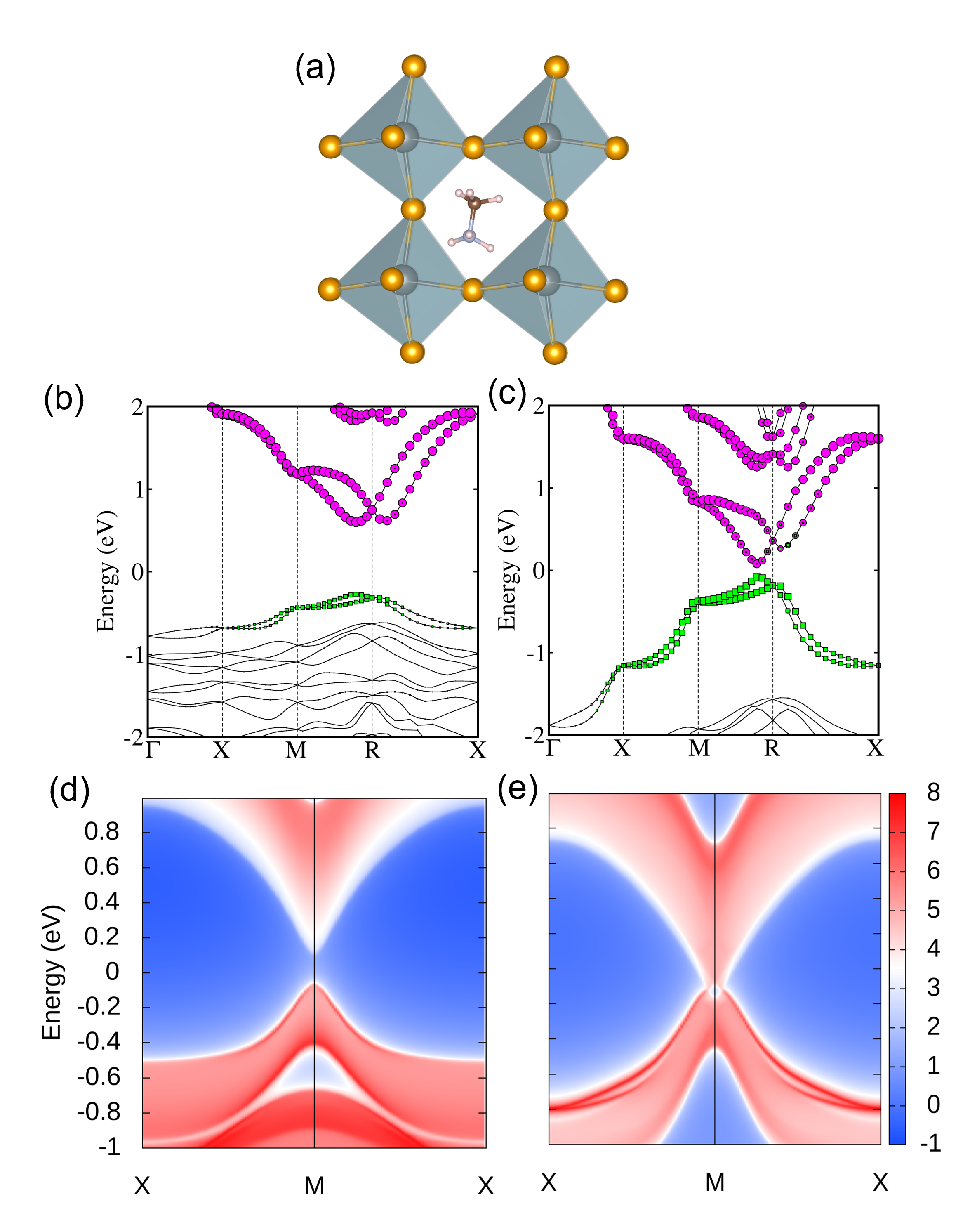}
	\caption{(a) Crystal structure of pseudocubic MAPbI$_3$. Bulk band structure (b) without compression with Pb-\textit{p} contribution at CBM and Pb-\textit{s} contribution at VBM, and (c) with compression $\gamma$= 0.73, showing band inversion at R point. Surface band structure (d) without, and (e) with compression $\gamma$= 0.73, showing a discontinuous tranistion from normal to topological state.}
	\label{3}
\end{figure*}

\newpage

{\textbf{Crystal structures and structural stability of Cs$_{0.5}$MA$_{0.5}$PbI$_3$, Cs$_{0.75}$MA$_{0.25}$PbI$_3$ and Cs$_{0.25}$MA$_{0.75}$PbI$_3$.}}\\

\begin{figure*}[htp]
	\includegraphics[width=0.7\textwidth]{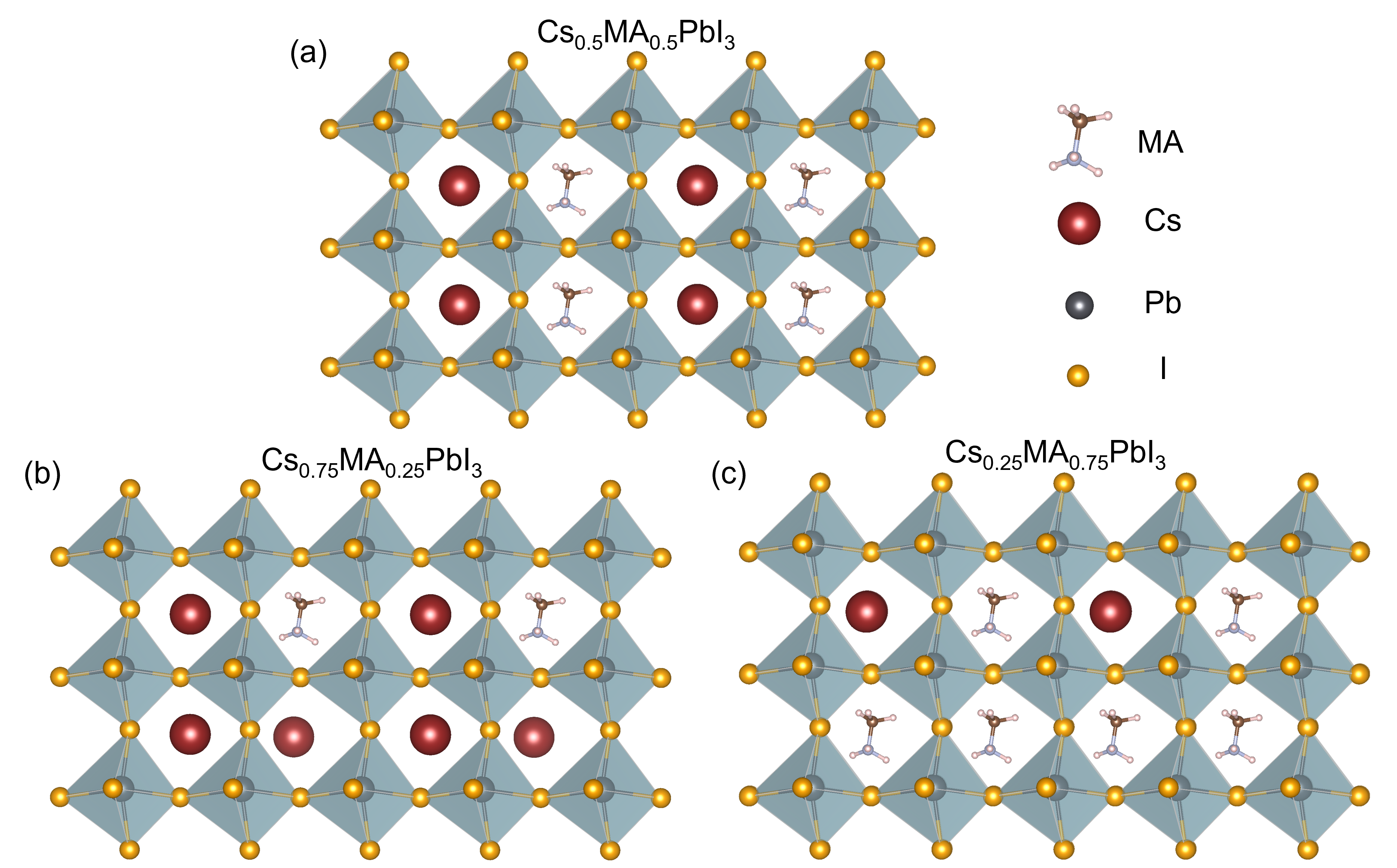}
	\centering
	\caption{Crytsal structure of (a) Cs$_{0.75}$MA$_{0.5}$PbI$_3$, (b) Cs$_{0.5}$MA$_{0.5}$PbI$_3$ and (c) Cs$_{0.25}$MA$_{0.75}$PbI$_3$.}
	\label{pic6}
\end{figure*}
\begin{figure*}[htp]
	\includegraphics[width=0.7\textwidth]{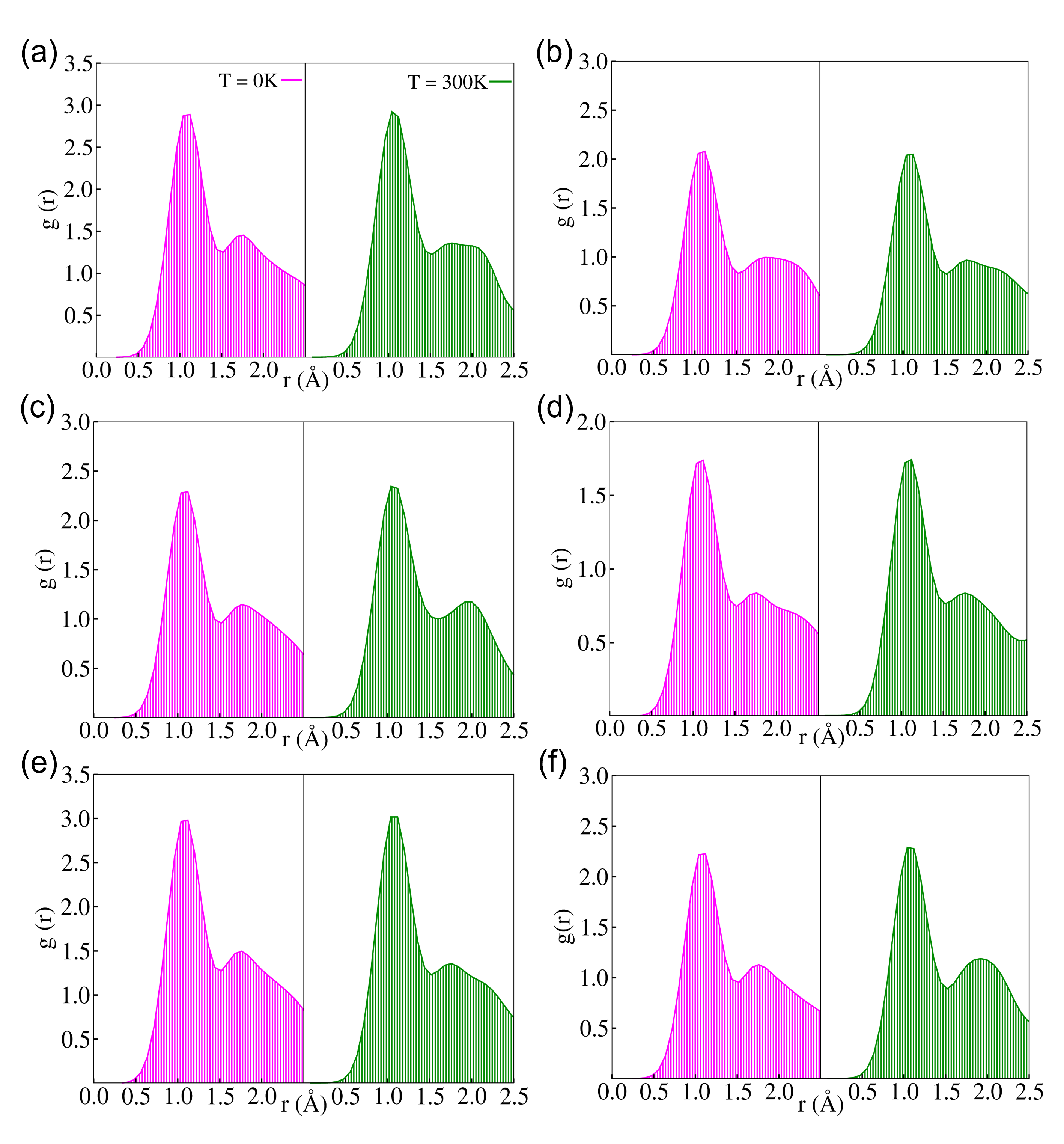}
	\caption{Radial distribution function at T = 0K and T = 300K for uncompressed and compressed ($\gamma$ = 0.74), (a), (b) Cs$_{0.5}$MA$_{0.5}$PbI$_3$, (c), (d) Cs$_{0.75}$MA$_{0.25}$PbI$_3$, (e) and (f) Cs$_{0.25}$MA$_{0.75}$PbI$_3$, respectively.}
	\label{4}
\end{figure*}

\newpage

{\textbf{TB parameters for Cs$_{0.5}$MA$_{0.5}$PbI$_3$. }}\\
\begin{table}[htbp]
	\caption {Interaction parameters and SOC strength for Cs$_{0.5}$MA$_{0.5}$PbI$_3$ from thirteen band hamiltonian at zero and critical compression strength in units of eV.}
	\begin{center}
		\begin{adjustbox}{width=0.8\textwidth} 
			\setlength\extrarowheight{+4pt}
			\begin{tabular}[c]{|c|c|c|c|c|c|c|c|c|c|c|c|} \hline		
				Type&$E_{Pb-s}$ & $E_{Pb-p}$ & $E_{I-p}$ &${t_{sp\sigma}^{Pb-I}}$   & ${t_{pp\sigma}^{Pb-I}}$   & ${t_{pp\pi}^{Pb-I}}$ & ${t_{ss\sigma}^{Pb-Pb}}$  & ${t_{sp\sigma}^{Pb-Pb}}$  & ${t_{pp\sigma}^{Pb-Pb}}$ & ${t_{pp\pi}^{Pb-Pb}}$ & $\lambda$ \\ \hline
				$\gamma$ = 1& -7 & 2  & -1.5 & 0.8 & -1  & 1 & -0.25 & 0.3 &  0.7& 0.01 & 0.5 \\ \hline
				$\gamma$ = 0.74 &-7 & 2.5  & -2 & 1 & -1.2  & 1.5 & -0.27 & 0.4 &  1.15 & 0.02 & 0.5     \\ \hline
			\end{tabular}
		\end{adjustbox}	
		\label{T1}
	\end{center}
\end{table}

\end{document}



\begin{abstract}
The centrosymmetric halide perovskites undergo a continuous phase transition from a normal insulator to a topological insulator at the critical value of strain. Contrarily, in noncentrosymmetric halide perovskites, this phase transition is discontinuous. The noncentrosymmetry does not stabilize the gapless state, causing a discontinuity in the bandgap. We have employed the density functional theory and Slater-Koster formalism-based tight-binding Hamiltonian studies to understand the evolution of band topology under the compressive strain in the halide perovskites. Our study shows that both cubic and pseudocubic FAPbI$_3$ undergo a Pb $\textit{s-p}$ band inversion at $\gamma$ (V/V$_0$) = 0.76 and 0.73, respectively. The cubic perovskite shows the surface state at $\overline{M}$, whereas, the pseudocubic structure shows two conducting states in the neighbourhood of $\overline{M}$, unlike the conventional topological insulator. The Pb-Pb second nearest neighbor interactions determine this topological phase transition. Alongside, we have modeled mixed cation halide perovskites Cs$_x$MA$_{1-x}$PbI$_3$ (\textit{x} = 0.25, 0.5 and 0.75) to study their topological properties. Cs$_{0.5}$MA$_{0.5}$PbI$_3$ shows non-trivial topology at $\gamma$ = 0.74. In addition, we have checked the structural stability of different strained configurations using ab \textit{initio} molecular dynamics at operational temperature. Their structural stability under compression strengthens the experimental relevance.


	\begin{tocentry}
		\begin{figure}[H]%
			\includegraphics[width=0.9\columnwidth,clip]{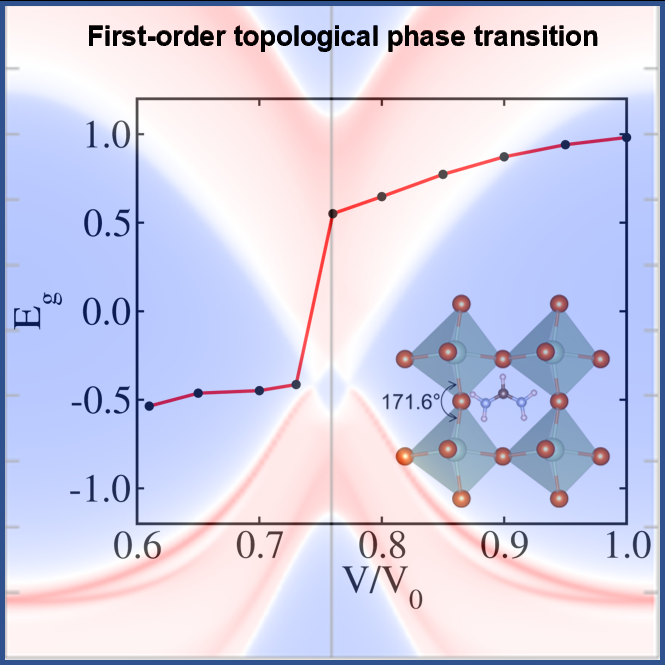}
		\end{figure}	
	\end{tocentry}
\end{abstract}

\section{Introduction}
Topological phases of matter such as topological insulators\cite{bansil2016colloquium}, Dirac semimetals\cite{wang2012dirac,wang2013three}, Weyl semimetals\cite{wan2011topological,xu2015discovery}, topological metals\cite{burkov2016topological}, and topological crystalline insulators\cite{fu2011topological} have occupied ample space in the research areas in the last two decades. Among these phases, topological insulators are one of the most promising materials for application in spintronics due to their conducting surface states within the bulk band gap\cite{hor2009p,xu2014enhanced,pal2015thermoelectric}. Numerous materials like Bi-based selenides and tellurides\cite{zhang2009topological,park2010robustness}, skutterudites\cite{yang2014dp}, Heusler compounds\cite{xiao2010half,feng2010half}, and Bi-based oxide perovskites (ABiO$_3$)\cite{yan2013large,li2015topological,khamari2018topologically} have been extensively explored for non-trivial topology. More recently, halide perovskites have also attracted attention for research in the area of band topology\cite{kore2020pressure}. These are the compounds with the formula ABX$_3$, where A is an inorganic or organic cation, B is a heavy element, and X is a halogen. These materials are well known for their high performance in the field of optoelectronics, photovoltaics, photocatalysts, and thermoelectrics\cite{jain2021theoretical,jain2020understanding,jain2023theoretical,bhumla2022vacancy}. Their endeavors can be attributed to the compositional and structural diversity that enables a wide array of functional properties. They possess a suitable orbital symmetry and are structurally flexible to external stimuli such as pressure, temperature, and chemical doping, which make them useful in the field of topology\cite{yang2012search,khamari2018topologically}. Furthermore, the molecular orbital picture of topologically non-trivial cubic ABiO$_3$, matches that of cubic halide perovskites (CsSnX$_3$)\cite{kashikar2018second}. However, unlike oxide perovskites, the $\textit{s}$-$\textit{p}$ inversion on inclusion of spin-orbit coupling (SOC) is not shown by the halide perovskites. Therefore, the family of halide perovskites is not known to show topologically non-trivial phases at ambient conditions. Nevertheless, if we tune certain aforementioned external parameters, a topological phase can be realized in halide perovskites\cite{kashikar2020defining}. It is consistent with the recent theoretical calculations that compressive strain can drive these materials into the topological insulator region\cite{kashikar2018second}. At a critical value of strain, a normal insulator to topological insulator (NI-TI) phase transition takes place. In materials preserving inversion symmetry, a continuous topological phase transition (TPT) take place\cite{jin2012topological}.  However, another type of phase transition, which is discontinuous in nature, has been observed in MAPbI$_3$ via. application of pressure. It has been shown that in pseudocubic perovskites like MAPbI$_3$, the inversion symmetry breaking has a profound effect on the nature of phase transition\cite{kore2020pressure}. The discontinuous phase transition has also been observed in Pb$_{1-x}$Sn$_x$Se and TlBiS$_{1-x}$Se$_x$\cite{krizman2018dirac,wojek2014band,jurivcic2017first,zeljkovic2015dirac}.
This is of key importance for device applications based on tuning the topological properties of the system by external means. Therefore, the realization of non-trivial phases under the effect of strain in pseudocubic perovskite will bring several families of compounds, beyond the intermetallic alloys, into the domain of research on band topology. This could be important in providing various heterostructure interfaces with multifunctional properties.\\
\indent Motivated by this idea, we have looked for the NI-TI phase transition in FAPbI$_3$ (FA = formamidinium) under the effect of compressive strain. It has been found that the nature of TPT is continuous for cubic FAPbI$_3$. But, when the centrosymmetry is broken in FAPbI$_3$, an inverted band gap opens without closing to zero. The absence of inversion symmetry brings a first-order quantum phase transition on application of strain. We have applied volumetric strain to trigger this non-trivial topology, by inducing a band inversion in the bulk band structure. After that, the surface band structure is obtained to see the origin of surface states with increasing strain. To explore the underlying quantum mechanical process, which occurs due to structural deformation caused by the application of strain, Slater-Koster tight-binding (SK-TB) model have been employed. We have also designed the thermodynamically stable mixed cation halide perovskites with the formula Cs$_x$MA$_{1-x}$PbI$_3$ (\textit{x} = 0.25, 0.5, and 0.75). The bulk and surface band structures are plotted at different values of strain to observe the TPT. Afterward, the structural stability of all the unstrained and strained configurations at an operational temperature of 300K has been confirmed to signify their practical usage.

\begin{figure*}[h]
	\includegraphics[width=0.85\textwidth]{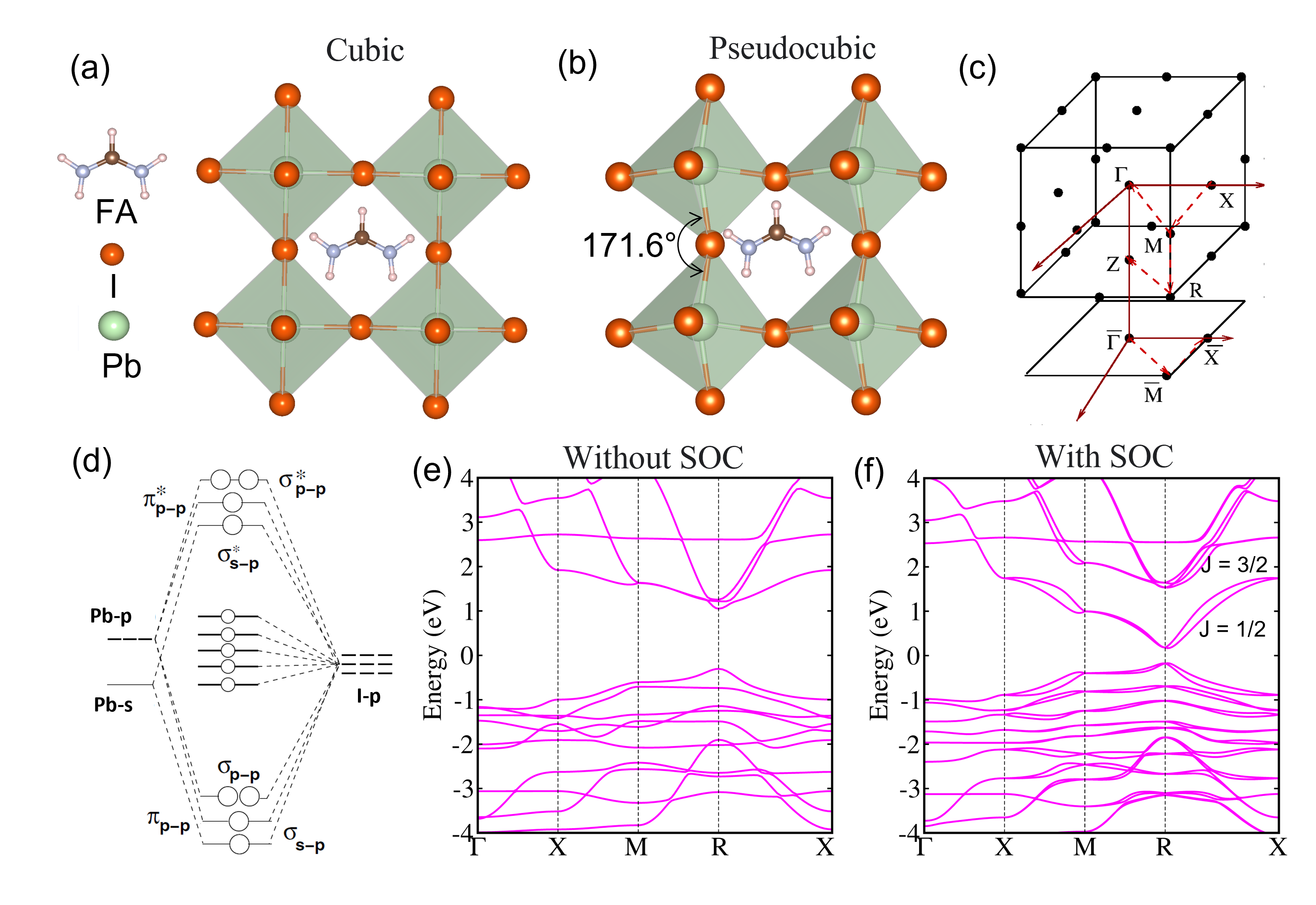}
	\centering
	\caption{Crystal structure of (a) cubic and (b) pseudocubic FAPbI$_3$. (c) 3D BZ along with the projected 2D BZ. (d) The molecular orbital picture for FAPbI$_3$. Bulk band structure of FAPbI$_3$ in the (e) absence of SOC and (f) presence of SOC.}
	\label{pic1}
\end{figure*}

FAPbI$_3$ exists in cubic phase with \textit{P}\textit{m}$\bar{3}$\textit{m} space group at 298 K with planar [H$_2$N-CH-NH$_2$]$^+$ cation surrounded by [PbI$_3$]$^-$ octahedra as shown in Figure \ref{pic1}(a). The lattice constant is taken as 6.32 \AA \:, which is in close agreement with the previous report\cite{weller2015cubic}. The structural relaxation of FAPbI$_3$ carried out using DFT methods leads to a distortion in the octahedral cage, causing an asymmetry in the structure\cite{banerjee2019rashba}. The crystal structure becomes pseudocubic with Pb-I-Pb angle $171.6^\circ$ as shown in Figure \ref{pic1}(b), which is similar to that of MAPbI$_3$\cite{kore2020pressure}. The bulk Brillouin zone (BZ) showing the high symmetry $\textbf{\textit{k}}$-points and the projected surface BZ is shown in Figure \ref{pic1}(c). The conduction bands near the Fermi surface are mainly contributed by the Pb-6\textit{p} orbitals, whereas the valence bands are mainly contributed by Pb-6\textit{s} and I-5\textit{p} orbitals\cite{wang2020structural}. The energy states of other elements in FA are mainly distributed in the deep energy levels, and are not involved in bonding. The Pb-\{\textit{s},\textit{p}\}--X-\textit{p} covalent interaction in the octahedron gives rise to four bonding and antibonding bands along with five nonbonding bands\cite{kashikar2021generic}. FA contributes one electron, Pb contributes four and I contribute five electrons per formula unit, hence, the valence electron count becomes 20. Therefore, the states up to $\sigma^*_{s-p}$ are filled and the Fermi level exists in between the lower-lying Pb-\textit{s} and upper-lying Pb-\textit{p} antibonding bands (see Figure \ref{pic1}(d)). The SOC arising due to the presence of heavy element affects the band structure near the Fermi level. Figure \ref{pic1}(e) shows the electronic band structure of FAPbI$_3$ in the absence of SOC. On inclusion of SOC, the \textbf{L.S} coupling lifts the degeneracy between the states \textbf{J} = $\frac{1}{2}$ and \textbf{J} = $\frac{3}{2}$ as shown in Figure \ref{pic1}(f). The strength of atomic SOC is the deterministic factor towards tailoring the non-trivial topological phases in ABX$_3$. However, the breaking down of the inversion symmetry causes both the atomic SOC and Rashba SOC to become deterministic in tailoring the electronic behavior of these noncentrosymmetric quantum materials\cite{frohna2018inversion}.
\begin{figure*}[h]
	\includegraphics[width=0.9\textwidth]{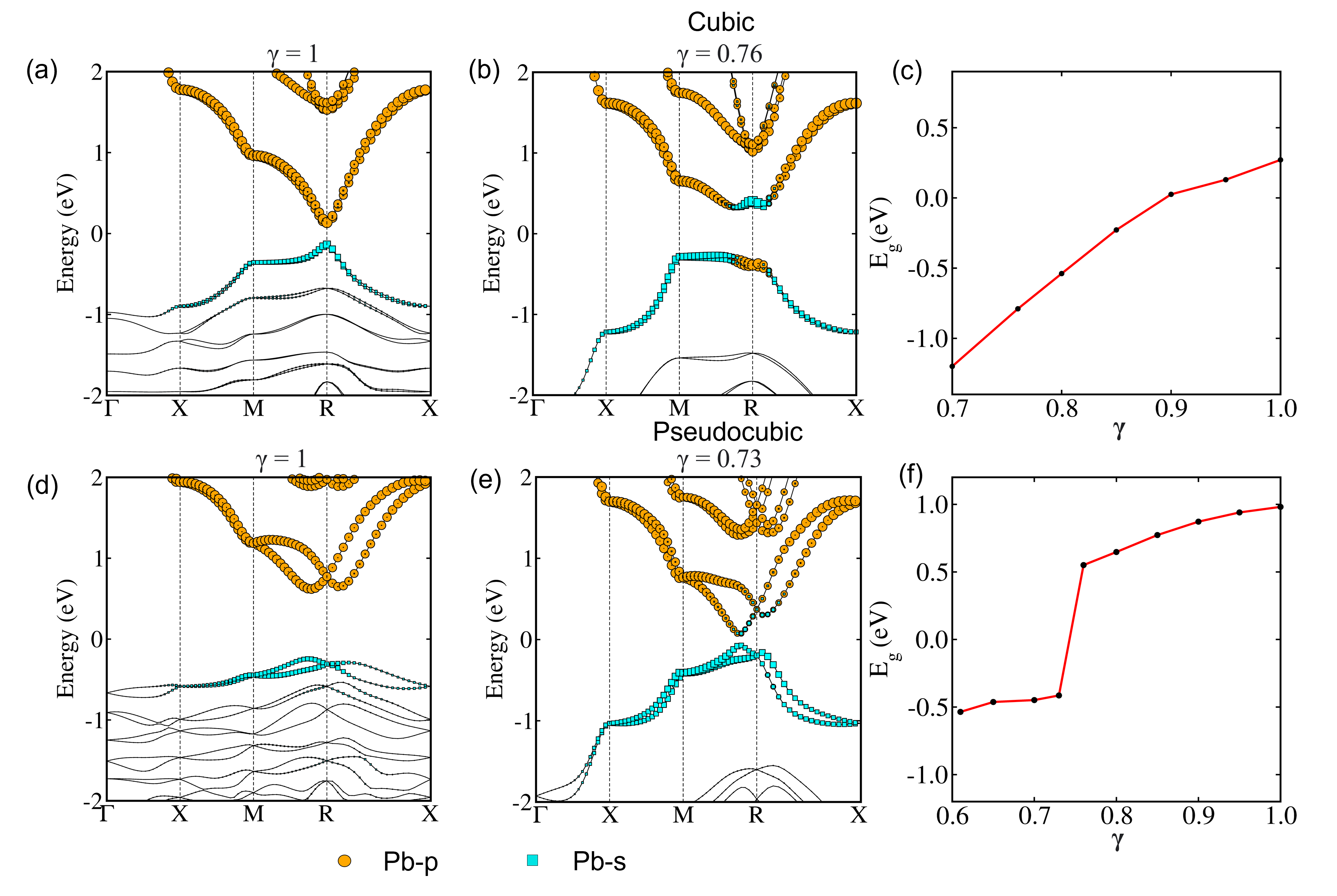}
	\centering
	\caption{Orbital resolved band structure in presence of SOC for cubic and pseudocubic FAPbI$_3$ (a), (d) without compression, and (b), (e) at critical compression $\gamma$ = 0.76, and $\gamma$ = 0.73, respectively. Band gap at different values of strain in (c) cubic FAPbI$_3$ showing its continuous nature, and (f) pseudocubic FAPbI$_3$ showing a discontinuity at the critical value of strain.}
	\label{pic2}
\end{figure*}

The experimentally reported band gap of FAPbI$_3$ is 1.48 eV\cite{aharon2015temperature}. The closest band gap of 1.56 eV is reproduced by G$_0$W$_0$@HSE06+SOC with $\alpha$=53\%, which is in agreement with the previous report\cite{basera2021capturing}. We have observed that except for the band gaps, other band features are similar irrespective of the choice of functional (see section I of supporting information (SI)). Therefore, we have used Perdew-Burke-Ernzerhof (PBE) functional for further calculations in view of its low computational cost.

The compressional strain is a natural external stimulus to vary the bond length while maintaining the symmetry of the lattice. Therefore, we have investigated the effect of strain on the evolution of band structure. In the case of cubic FAPbI$_3$, at a critical compression $\gamma$ (V/V$_0$) = 0.76, the inversion between the Pb-$\textit{s}$ and Pb-$\textit{p}$ orbital takes place. This $\textit{s-p}$ inversion takes place at R point and gives the initial indication of the existence of non-trivial topology under the effect of strain. The orbital projected band structure with the orbital contribution proportional to the thickness of the symbols is shown in Figures \ref{pic2}(a) and (b). The continuous TPT takes place in three steps, i) shrinking ii) closing iii) reemergence of the inverted band gap, and is shown in the Figure \ref{pic2}(c). However, if the inversion symmetry is not preserved in a material then the nature of phase transition changes. In the case of noncentrosymmetric FAPbI$_3$, as we keep on increasing the strain the band gap decreases initially. But, after reaching a certain minimum value, a new negative band gap emerges with VBM and CBM inverting their orbital character. In this case, the value of this critical strain $\gamma$ is 0.73 (see  Figures \ref{pic2}(d) and (e)). This flip in the character of band structure at the critical value indicates the change in the topological nature of the material. This type of phase transition in which the gapless state is not stabilized, resulting in discontinuity in the band gap as shown in Figure \ref{pic2}(f), is termed a discontinuous TPT. It is also called a first-order phase transition. The distortion caused in the octahedral cage of PbI due to structural optimization is responsible for the discontinuous phase transition. Subsequently, the structural stability of cubic and pseudocubic FAPbI$_3$ along with the strained configurations is confirmed using ab \textit{initio} molecular dynamics (AIMD). The radial distribution function (g(r)) plots at T = 0 K and T = 300 K are shown in section II of SI.


SK-TB Hamiltonian studies are useful in providing information regarding the underlying quantum mechanical process responsible for phase transition due to structural deformation\cite{boyer2016symmetry}. The TB model Hamiltonian in the second quantized notation is written as\cite{do2022generalized,kashikar2021generic}
\begin{equation}
H = \sum \limits_{i,\alpha} \epsilon_{i\alpha} c_{i\alpha}^\dagger c_{i\alpha} + \sum \limits_{ij:\alpha \beta}  t_{i\alpha j\beta} (c_{i\alpha}^\dagger c_{j\beta} + h.c.) + \lambda\textbf{L.S} + H_{R}.
\end{equation}
Here, i,j and $\alpha$, $\beta$ represents site and orbital indices, respectively. $\epsilon_{i\alpha} $ is the onsite energy term and $t_{i\alpha j\beta}$ gives the hopping term. In addition to the nearest neighbor Pb-I interactions, the second neighbor Pb-Pb interactions are also included in the Hamiltonian. The third term includes SOC, where $\lambda$ gives the SOC strength. \textit{H$_R$} is included when the noncentrosymmetry is present in the crystal structure and is called Rashba SOC. \textit{H$_R$} = --$\frac{\mu_B}{2mc}$$\vec\sigma$.($\vec{E}$$\times$$\vec{p}$), where \textit{$\vec{\textbf{E}}$} is the electric field experienced by the electron and \textbf{\textit{$\vec\sigma$}} is the pauli matrix vector and \textit{{$\vec{\textbf{p}}$}} represents the momenta\cite{sheoran2022rashba,sheoran2022emergence}. The basis set for the TB Hamiltonian contains one Pb-$\textit{s}$, three Pb-$\textit{p}$ and nine I-$\textit{p}$ orbitals. The rest orbitals are neglected as they do not participate in the formation of the topological invariant states. The spin independent Hamiltonian with the basis set in order $\bra{s^{Pb}}$, $\bra{p_{x}^{Pb}}$, $\bra{p_{y}^{Pb}}$, $\bra{p_{z}^{Pb}}$, $\bra{p_{x}^{I1}}$, $\bra{p_{y}^{I1}}$, $\bra{p_{z}^{I1}}$,$\bra{p_{x}^{I2}}$, $\bra{p_{y}^{I2}}$, $\bra{p_{z}^{I2}}$, $\bra{p_{x}^{I3}}$, $\bra{p_{y}^{I3}}$, $\bra{p_{z}^{I3}}$, takes the form
$$
H =
\begin{pmatrix}
M_{4\times4}^{Pb-Pb} & M_{4\times9}^{Pb-I}\\
(M_{4\times9}^{Pb-I})^\dagger & M_{9\times9}^{I-I} \\
\end{pmatrix}
.
$$
The individual blocks of this matrix are as follows
$$
M_{4\times4}^{Pb-Pb} =
\begin{pmatrix}
\epsilon_s + g_1 & 2it_{sp\sigma}^{Pb-Pb} Sin(k_xa) & 2it_{sp\sigma}^{Pb-Pb} Sin(k_ya) & 2it_{sp\sigma}^{Pb-Pb} Sin(k_za) \\
-2it_{sp\sigma}^{Pb-Pb} Sin(k_ya) & \epsilon_{p1} + g_2 & 0 & 0\\
-2it_{sp\sigma}^{Pb-Pb} Sin(k_xa) & 0 & \epsilon_{p1} + g_3 & 0\\
-2it_{sp\sigma}^{Pb-Pb} Sin(k_za) & 0 & 0 & \epsilon_{p1} + g_4\\
\end{pmatrix}
$$

$$
M_{4\times9}^{Pb-I} =
\begin{pmatrix}
t_{sp\sigma}^{Pb-I} S_x & 0 & 0 & 0 & t_{sp\sigma}^{Pb-I} S_y & 0 & 0 & 0 & t_{sp\sigma}^{Pb-I} S_z\\
t_{pp\sigma}^{Pb-I} C_x &  0 & 0 & t_{pp\pi}^{Pb-I} C_y & 0 & 0 & t_{pp\pi}^{Pb-I} C_z  & 0 & 0\\
0 & t_{pp\pi}^{Pb-I} C_x & 0 & 0 & t_{pp\pi}^{Pb-I} C_y & 0 & 0 & t_{pp\pi}^{Pb-I} C_z  & 0\\
0 & 0 & t_{pp\pi}^{Pb-I} C_x & 0 & 0 & t_{pp\pi}^{Pb-I} C_y & 0 & 0 & t_{pp\sigma}^{Pb-I} C_z \\
\end{pmatrix}
$$

$$
M_{9\times9}^{I-I} =
\begin{pmatrix}
\epsilon_{p2} & 0 & 0 & 0 & 0 & 0 & 0 & 0 & 0 \\
0 & \epsilon_{p2} & 0 & 0 & 0 & 0 & 0 & 0 & 0 \\
0 & 0 & \epsilon_{p2} & 0 & 0 & 0 & 0 & 0 & 0 \\
0 & 0 & 0 & \epsilon_{p2} & 0 & 0 & 0 & 0 & 0 \\
0 & 0 & 0 & 0 & \epsilon_{p2} & 0 & 0 & 0 & 0 \\
0 & 0 & 0 & 0 & 0 & \epsilon_{p2} & 0 & 0 & 0 \\
0 & 0 & 0 & 0 & 0 & 0 & \epsilon_{p2} & 0 & 0 \\
0 & 0 & 0 & 0 & 0 & 0 & 0 & \epsilon_{p2} & 0 \\
0 & 0 & 0 & 0 & 0 & 0 & 0 & 0 & \epsilon_{p2} \\
\end{pmatrix}
.
$$
Here $\epsilon_{s}$ ,$\epsilon_{p1}$ , and $\epsilon_{p2}$ are the onsite energies of Pb-\textit{s}, Pb-\textit{p} and I-\textit{p} orbitals, respectively. The terms  $C_z$ and $S_z$ are the representations
for 2$\cos(k_za/2)$ and 2$\textit{i}$$\sin(k_za/2)$, and  $g_i$ (i = 1,2,3,4) are the \textbf{\textit{k}}-dependent hopping integrals arising from the Pb-Pb second neighbor interactions given by

\begin{align*}
g_1 = 2t_{ss}^{Pb-Pb} [\cos(k_xa)+\cos(k_ya)+\cos(k_za)], \\
g_2 = 2t_{pp\sigma}^{Pb-Pb}\cos(k_xa) + 2t_{pp\pi}^{Pb-Pb}[\cos(k_ya)+\cos(k_za)], \\
g_3 = 2t_{pp\sigma}^{Pb-Pb}\cos(k_ya) + 2t_{pp\pi}^{Pb-Pb}[\cos(k_xa)+\cos(k_za)], \\
g_4 = 2t_{pp\sigma}^{Pb-Pb}\cos(k_za) + 2t_{pp\pi}^{Pb-Pb}[\cos(k_xa)+\cos(k_ya)]. \\
\end{align*}

The SOC component of the Hamiltonian is obtained by applying $\lambda$\textbf{L.S} on the Pb-\textit{p} states, as Pb being heavy is responsible for the SOC. We have obtained the matrix elements for the SOC Hamiltonian with the basis set in the order:
$\ket{p_{x}^{Pb} \uparrow}$,  $\ket{p_{y}^{Pb} \uparrow}$,  $\ket{p_{z}^{Pb} \uparrow}$,  $\ket{p_{x}^{Pb} \downarrow}$,  $\ket{p_{y}^{Pb} \downarrow}$,  $\ket{p_{z}^{Pb} \downarrow}$

$$
H_{SOC} = \lambda
\begin{pmatrix} 
0 & -i & 0 & 0 & 0 & 1\\
i & 0 & 0 & 0 & 0 & -i \\
0 & 0 & 0 & 1 & -i & 0\\
0 & 0 & 1 & 0  & i & 0 \\
0 & 0 & i & -i & 0 & 0\\
1 & i & 0 & 0 & 0 & 0\\
\end{pmatrix}
.
$$
\begin{table}[htbp]
	\caption {Interaction parameters and SOC strength for FAPbI$_3$ from thirteen band hamiltonian at zero and critical compression strength in units of eV.}
	\begin{center}
		\begin{adjustbox}{width=0.8\textwidth} 
			\setlength\extrarowheight{+4pt}
			\begin{tabular}[c]{|c|c|c|c|c|c|c|c|c|c|c|c|} \hline		
				Type&$E_{Pb-s}$ & $E_{Pb-p}$ & $E_{I-p}$ &${t_{sp\sigma}^{Pb-I}}$   & ${t_{pp\sigma}^{Pb-I}}$   & ${t_{pp\pi}^{Pb-I}}$ & ${t_{ss\sigma}^{Pb-Pb}}$  & ${t_{sp\sigma}^{Pb-Pb}}$  & ${t_{pp\sigma}^{Pb-Pb}}$ & ${t_{pp\pi}^{Pb-Pb}}$ & $\lambda$ \\ \hline
				Cubic ($\gamma$ = 1) & -1  & 2.5 & -2 & 0.15  & 0.4 & 0.3 & -0.1 &  0.3& 0.4 & 0.1 & 0.5  \\ \hline 
				Cubic ($\gamma$ = 0.76) & -1  & 2.5 & -2 & 0.1  & 0.7 & 0.5 & -0.1 &  0.3& 0.76 & 0.17 & 0.5 \\ \hline 
				Pseudocubic ($\gamma$ = 1) & -0.8  & 3 & -1 & 0.15  & 0.4 & 0.3 & -0.07 &  0.45& 0.5 & 0.4 & 0.4 \\ \hline
				Pseudocubic ($\gamma$ = 0.73) & -1  & 3 & -2 & 0.1  & 0.5 & 0.5 & -0.11 &  0.5& 0.95 & 0.07 & 0.4     \\ \hline
			\end{tabular}
		\end{adjustbox}	
		\label{T1}
	\end{center}
\end{table}

Fitting this thirteen band model with the DFT bands, we get the effective onsite and hopping terms. The parameters obtained at no strain and critical strain are provided in Table \ref{T1}. In the case of FAPbI$_3$, the bands forming the VBM and CBM are primerly composed of Pb-\{\textit{s},\textit{p}\} orbitals. Also, the band inversion is taking place between Pb-\textit{s}, and Pb-\textit{p} orbitals. Therefore, a minimal basis Hamiltonian containing only essential interactions have been developed. The second neighbor Pb-Pb interactions has been considered in formulating the four band model. The SOC incorporated four band Hamiltonian is written as

$$
\textbf{H} =
\begin{pmatrix} 
H_{\uparrow\uparrow} & H_{\uparrow\downarrow} \\
H^{\dagger}_{\downarrow\uparrow} & H_{\downarrow\downarrow}  \\
\end{pmatrix}
$$
where
$$
H_{\uparrow\downarrow} =
\begin{pmatrix} 
0 & 0 & 0 & 0 \\
0 & 0 & 0 & \lambda \\
0 & 0 & 0 & -i\lambda\\
0 & 0 & \lambda & -i\lambda \\
\end{pmatrix}
$$

$$
H_{\uparrow\uparrow} =
\begin{pmatrix}
\epsilon_s + g_1 & 2it_{sp\sigma}^{Pb-Pb} Sin(k_xa) & 2it_{sp\sigma}^{Pb-Pb} Sin(k_ya) & 2it_{sp\sigma}^{Pb-Pb} Sin(k_za) \\
-2it_{sp\sigma}^{Pb-Pb} Sin(k_ya) & \epsilon_{p1} + g_2 & 0 & 0\\
-2it_{sp\sigma}^{Pb-Pb} Sin(k_xa) & 0 & \epsilon_{p1} + g_3 & 0\\
-2it_{sp\sigma}^{Pb-Pb} Sin(k_za) & 0 & 0 & \epsilon_{p1} + g_4\\
\end{pmatrix}
$$
\begin{table}[htbp]
	\caption {Interaction parameters and SOC strength for FAPbI$_3$ from four band hamiltonian at zero and critical compression strength in units of eV.}
	\begin{center}
		\begin{tabular}[c]{|c|c|c|c|c|c|c|c|} \hline		
			Type & $E_{Pb-s}$ & $E_{Pb-p}$  & ${t_{ss\sigma}^{Pb-Pb}}$  & ${t_{sp\sigma}^{Pb-Pb}}$  & ${t_{pp\sigma}^{Pb-Pb}}$ & ${t_{pp\pi}^{Pb-Pb}}$ & $\lambda$ \\ \hline
			Cubic ($\gamma$ = 1) & -0.8  & 2.5 & -0.09 & 0.3  & 0.4 & 0.1 & 0.5       \\ \hline
			Cubic ($\gamma$ = 0.76) &-1.2  & 2.5 & -0.19 & 0.44  & 0.75 & 0.16 & 0.5       \\ \hline
			Pseudocubic ($\gamma$ = 1) & -0.8  & 2.5 & -0.1 & 0.25  & 0.2 & 0.06 & 0.5       \\ \hline
			Pseudocubic ($\gamma$ = 0.73) & -1  & 3 & -0.12 & 0.4  & 0.8 & 0.1  & 0.5       \\ \hline
		\end{tabular}
		\label{T2}
	\end{center}
\end{table}

The interaction parameters for both cubic and pseudocubic FAPbI$_3$, from the four band model fitting are provided in Table \ref{T2}. Magnitude of all the parameters increase with rise in strain. With the increase in compression, the bond length decreases and hence the strength of interactions increases. In comparison to Pb-Pb ss$\sigma$, sp$\sigma$ and pp$\pi$, the pp$\sigma$ interactions are most sensitive to the strain. For cubic structure the pp$\sigma$ interaction rises from 0.4 to 0.75, while for pseudocubic, it rises from 0.2 to 0.7, upon the application of respective critical strain.  We have also observed that $\lambda$ does not change with compression, therefore, SOC is insensitive to the strain. From the electronic band structure calculations, it is evident that the universality of the band structure arises from the nearest-neighbor Pb-\{\textit{s},\textit{p}\}--X-\textit{p} covalent interactions. However, the TB model elucidates that the second neighbor Pb-\{\textit{s},\textit{p}\}--Pb-\{\textit{s},\textit{p}\} interactions act as a driving force for band gap engineering and non-trivial phases. This type of TB model is universal, as it is independent of the choice of functional adopted\cite{kashikar2021generic}.

\begin{figure*}[h]
	\includegraphics[width=0.85\textwidth]{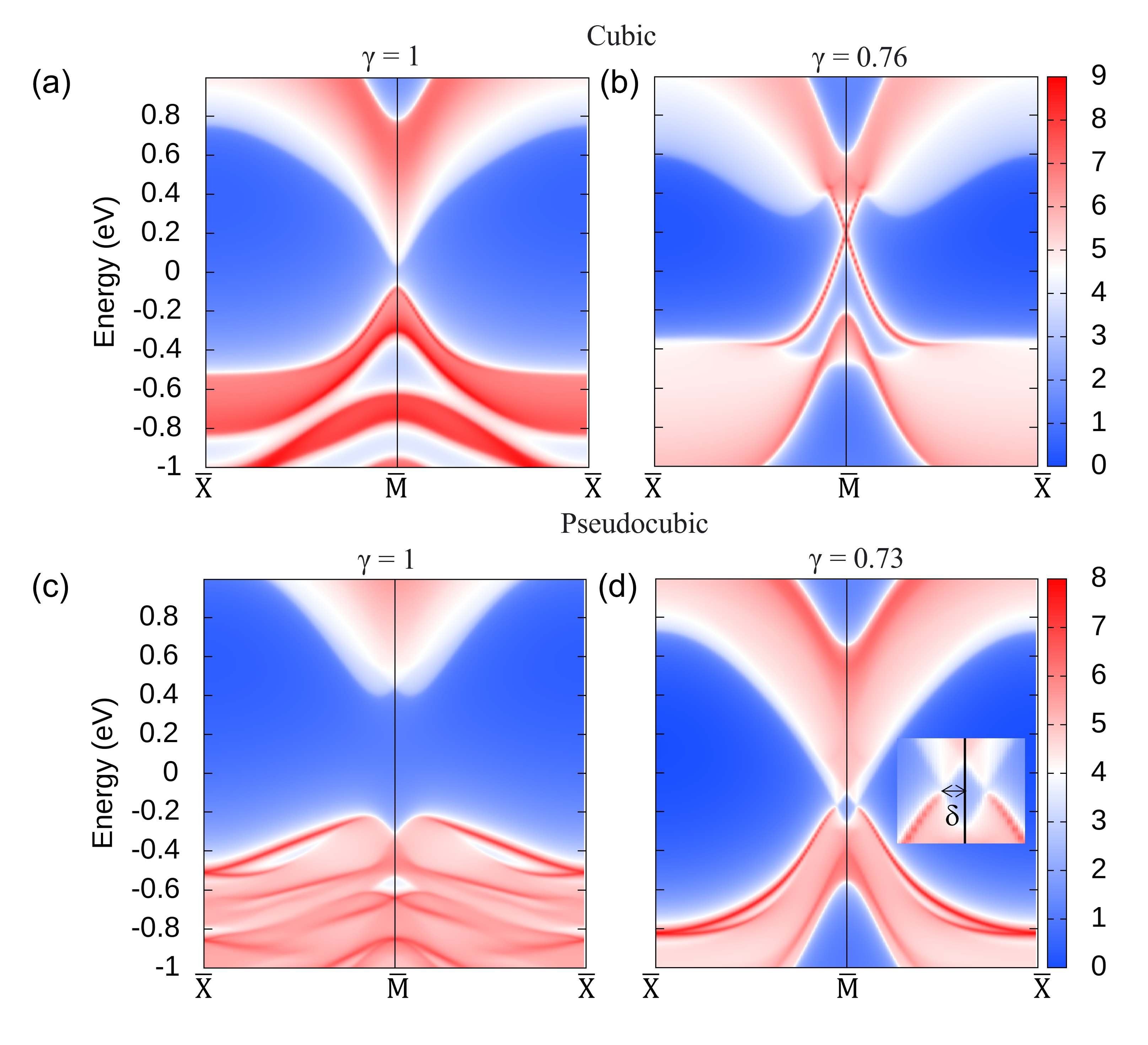}
	\centering
	\caption{Surface band structure in presence of SOC for cubic FAPbI$_3$ (a) without compression, (b) at critical compression $\gamma$ = 0.76, and pseudocubic FAPbI$_3$ (c) without compression, (d) at critical compression $\gamma$ = 0.73.}
	\label{pic4}
\end{figure*}

The study of bulk electronic structure has enabled us to identify the possibility of non-trivial topology and phase transition in halide perovskites. However, the practical observation of these states comes from the calculation of surface states\cite{phutela2022exploring}. 
The DFT based calculations on the slab structure require a huge amount of computational time and memory to analyze these properties. So, we have used the maximally localized wannier function (MLWF) based TB approach to construct a semi-infinite slab along (001) direction, minimizing the total spread of MLWF. As expected, there is an absence of the surface states in FAPbI$_3$ at ambient conditions (see Figures \ref{pic4}(a) and (c)). However, as the critical value of strain i.e., $\gamma$ = 0.76 is applied to cubic FAPbI$_3$, the conducting states are formed at $\overline{M}$ point, similar to that of a conventional topological insulator. The invariant surface states, exhibited by cubic FAPbI$_3$ yields two interpenetrating Dirac states. In this case, the Kramer pair i.e., doubly degenerate dirac cones, each one arising from the VB and CB coincides at $\overline{M}$ point (see Figure \ref{pic4}(b)). But when the inversion symmetry is broken, this degeneracy is lifted and as a result one cone shifts towards the VB and the other towards the CB\cite{kore2020pressure}. Therefore, we see two Dirac states  at $\overline{M}$ $\pm$ $\delta$, along the path X--$\overline{M}$--X, as shown in Figure \ref {pic4}(d). The resulting surface states illustrate the presence of discontinuous nature of NI-TI TPT.

\begin{figure*}[h]
	\includegraphics[width=0.85\textwidth]{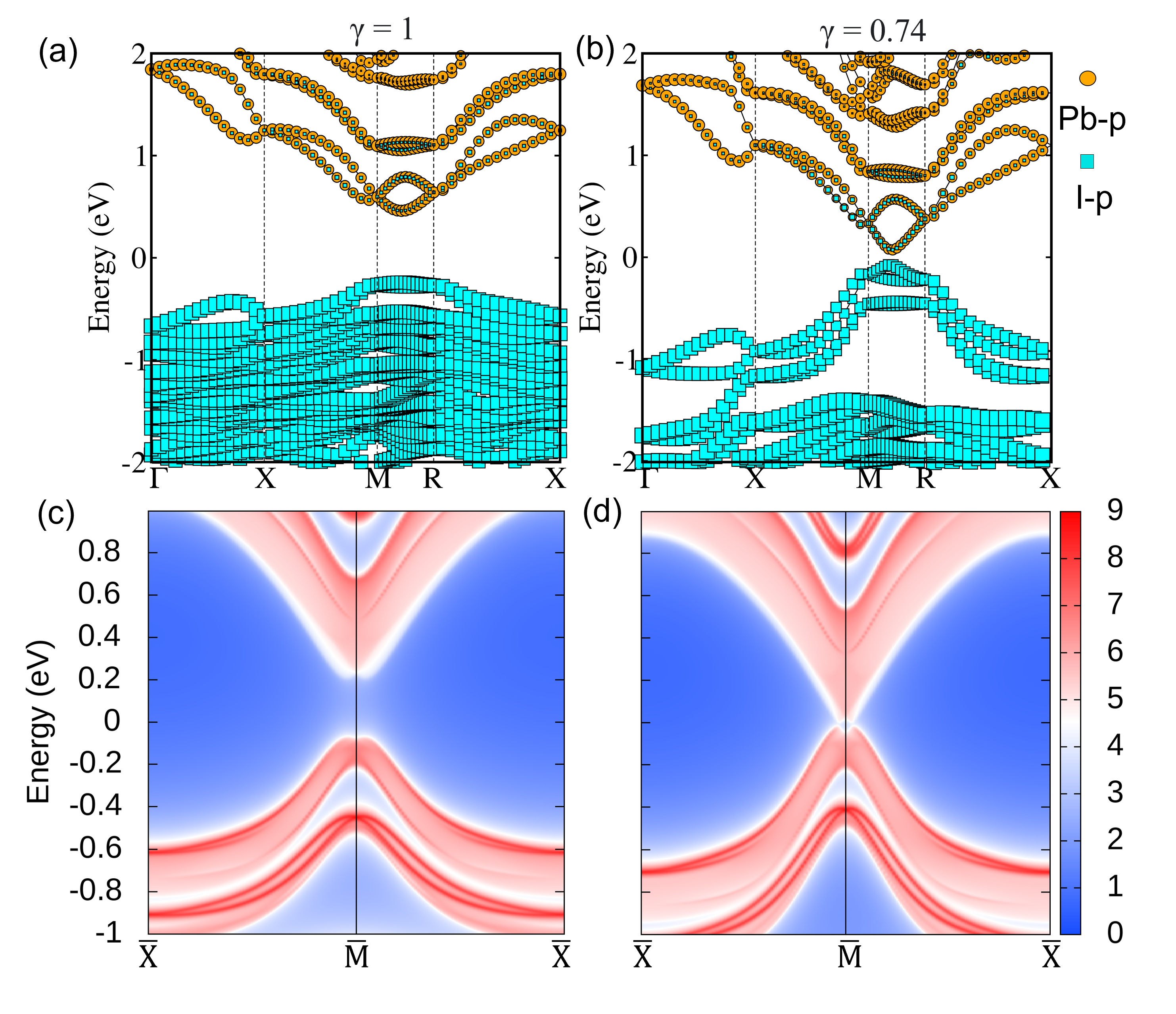}
	\centering
	\caption{Orbital resolved band structure in presence of SOC for Cs$_{0.5}$MA$_{0.5}$PbI$_3$ (a) without compression, (b) at critical compression $\gamma$ = 0.74. Surface band structure for Cs$_{0.5}$MA$_{0.5}$PbI$_3$ (a) without compression, (b) at critical compression $\gamma$ = 0.74.}
	\label{pic7}
\end{figure*}

We have studied the TPT in both inorganic (CsPbI$_3$) and organic (MAPbI$_3$, FAPbI$_3$) halide perovskites. CsPbI$_3$ show continuous TPT at $\gamma$ = 0.76 because of its centrosymmetric nature, whereas MAPbI$_3$ and FAPbI$_3$ show similar behaviour (see section III and IV of SI for CsPbI$_3$ and MAPbI$_3$, respectively). It has been found that by partially replacing MA$^+$ with Cs$^+$, the stability can be enhanced in ambient environment\cite{ma2018substituting}. These mixed cation/anion perovskites have attracted considerable attention due to their higher stability\cite{gao2018robust}. Thus, we have designed the mixed cation halide perovskites to study the nature of TPT in them. For this, three configurations i.e., Cs$_{0.75}$MA$_{0.5}$PbI$_3$, Cs$_{0.5}$MA$_{0.5}$PbI$_3$, Cs$_{0.25}$MA$_{0.75}$PbI$_3$, have been obtained by replacing MA with Cs in the lattice of MAPbI$_3$ (see section V of SI for crystal structures). Firstly, we have calculated the thermodynamic stability of the mixed cation halide perovkites i.e., Cs$_{0.75}$MA$_{0.5}$PbI$_3$, Cs$_{0.5}$MA$_{0.5}$PbI$_3$, Cs$_{0.25}$MA$_{0.75}$PbI$_3$. The precursors considered for the calculation of formation energy (\textit{E$_f$}) of these perovskites are MAPbI$_3$, CsI and PbI$_2$. The associated expression is\\
\begin{equation}
E_{f}(x,y) = E_{sys} - E(MA_4Pb_4I_{12}) - xE(CsI) - yE(PbI_2)
\end{equation}
Here, coefficients \textit{x} and \textit{y} of each term on the right hand side are chosen such that they stoichiometrically balance the equation. 1$\times$2$\times$2 supercell of MAPbI$_3$ has been used. \textit{E$_f$(x,y)} for Cs$_{0.75}$MA$_{0.25}$PbI$_3$, Cs$_{0.5}$MA$_{0.5}$PbI$_3$, Cs$_{0.25}$MA$_{0.75}$PbI$_3$ is -0.78 eV, -0.69 eV and -0.65 eV, respectively. The negative formation energy shows that all the configurations are thermodynamically stable. We have also checked the structural stability of these configurations using AIMD simulations (see section V of SI). After this, we have obtained the band structure for all the configurations with and without SOC. When no strain is applied in Cs$_{0.5}$MA$_{0.5}$PbI$_3$, the VBM is composed of I-\textit{p} orbitals, and CBM of Pb-\textit{p} orbitals (see Figure \ref{pic7}(a)). The inclusion of SOC does not depict any traits of non-trivial topological nature. Subsequently, we have applied the compressional volumetric strain along with the SOC.  At the critical strain value of $\gamma$ = 0.74, the appearance of a negative band gap is observed with the inverted orbital character as seen in Figure \ref{pic7}(b). This suggests the possibility of TPT in Cs$_{0.5}$MA$_{0.5}$PbI$_3$. Whereas, in Cs$_{0.75}$MA$_{0.25}$PbI$_3$ and Cs$_{0.25}$MA$_{0.75}$PbI$_3$, no such parity inversion is observed under the effect of compression. This may be attributed to the difference in the ionic radii of Cs and MA cations, which is 1.88 \AA \: and 2.117 \AA \:, respectively. As the band inversion involves \textit{p} orbitals of both I and Pb, therefore, a four band TB model will not suffice in predicting the nature of TPT. Thus, using the thirteen band model containing Pb-\{\textit{s},\textit{p}\} and X\textit-{p} orbitals, we have calculated the onsite energies and hopping integrals (see section VI of SI). All the interaction parameters increase on application of strain, but the Pb-I pp$\pi$ interations rising from 1 to 1.5 are most affected by the same. After this, we have plotted the surface state band structure for Cs$_{0.5}$MA$_{0.5}$PbI$_3$ at $\gamma$ = 0 and $\gamma$ = 0.74. As seen in the Figures \ref{pic7}(b) and (c), the origin of surface states with strain confirms the TPT at $\gamma$ = 0.74. 

The strain free lead halide perovskites are trivial band insulators, however, a TPT can take place via. strain engineering. The presence of inversion symmetry plays a vital role in the electronic properties of the halide perovskites. The first-principles calculations suggest that FAPbI$_3$ becomes noncentrosymmetric on structural optimization. The centrosymmetric FAPbI$_3$ shows a continuous phase transition on the application of strain. The inversion symmetry breaking changes the nature of this quantum phase transition. At a critical value of $\gamma$ = 0.73, pseudocubic FAPbI$_3$ undergoes a first-order TPT. The discontinuity is evident from the band gaps calculated at various values of strain, and the nature of surface states. This study also shows that although thirteen orbitals participate in the chemical bonding, a four band TB model is sufficient to capture energy dispersion near the Fermi level. The Pb-Pb second neighbor interactions determine the topology of this class of materials. The Pb-Pb pp$\sigma$ interactions are the most affected by the application of strain. Hence, these interactions are the driving forces in determining the phase transition. Furthermore, the mixed cation perovskites Cs$_{0.75}$MA$_{0.25}$PbI$_3$, Cs$_{0.5}$MA$_{0.5}$PbI$_3$ and Cs$_{0.25}$MA$_{0.75}$PbI$_3$ are thermodynamically and structurally stable. Cs$_{0.5}$MA$_{0.5}$PbI$_3$ shows TPT at a compression of $\gamma$ = 0.74, which is shown by band inversion and the nature of surface states. The structural stability of these perovskites under strain strengthens that these phase transitions can be realized experimentally.

\section{Compulational Methods}
The calculations are performed using DFT\cite{hohenberg1964inhomogeneous,kohn1965self} with the Projected Augmented Wave (PAW)\cite{kresse1999ultrasoft,blochl1994projector} method implemented in Vienna \textit{Ab initio} Simulation Package (VASP)\cite{kresse1996efficient} code.  The self-consistent relativistic calculations are performed using the Generalized Gradient Approximation (GGA) of PBE\cite{perdew1996generalized}. The structural parameters are fully optimized until the Hellmann-Feynman forces are smaller than 1 meV/\AA. The cutoff energy of 520 eV is used for the plane wave basis set such that the total energy calculations are converged within 10$^{-5}$ eV. A $\Gamma$-centered 4$\times$4$\times$4 \textit{k}-grid is used to sample the irreducible BZ of cubic phase with the \textit{P}\textit{m}$\bar{3}$\textit{m} space group. Throughout the calculations, we have included SOC in order to consider the relativistic effects due to the presence of heavy element in the systems. We have used many-body perturbation method G$_0$W$_0$@HSE06+SOC for the better estimation of the band gap\cite{hedin1965new}. For this, the number of bands is set to six times the number of occupied bands. After this, to investigate the topological phases, the first-principles calculations were first performed using fully relativistic norm-conserving pseudopotentials as implemented in the QUANTUM ESPRESSO code\cite{giannozzi2009quantum}. The results of these DFT calculations are then fed as input to WANNIER90\cite{mostofi2008wannier90} for constructing a TB model based on Maximally Localized Wannier Functions (MLWFs) with \textit{p} orbitals of I and \textit{sp} orbitals Pb. The surface states are then calculated using the  Green’s function through an interactive method as implemented in the Wannier-TOOLS package\cite{wu2018wanniertools}.  A TB model using Slater and Koster approach is constructed in the two-center approximation using the package Tight Binding Studio\cite{nakhaee2020tight}. A nonlinear fitting algorithm is used to find the best fit entries for both Hamiltonian and overlap matrices to reproduce the first-principles data. Finally, the structural stability of different configurations is checked at higher temperature using AIMD by running a 5 ps long simulation with NVT ensemble (Nose–Hoover thermostat).  



	


\begin{acknowledgement}
	A.P. acknowledges IIT Delhi for the senior research fellowship. S.S. acknowledges CSIR, India, for the senior research	fellowship [grant no. 09/086(1432)/2019-EMR-I]. S.B. acknowledges financial support from SERB under a core research grant (grant no. CRG/2019/000647) to set up his High Performance Computing (HPC) facility ‘‘Veena’’ at IIT Delhi for computational resources.
\end{acknowledgement}
\begin{suppinfo}
Band profile of FAPbI$_3$ using PBE+SOC and G$_0$W$_0$@HSE06+SOC $\epsilon_{xc}$ functionals; Structural stability of cubic and pseudocubic FAPbI$_3$; Continous TPT in cubic CsPbI$_3$; TPT in cubic and pseudocubic MAPbI$_3$; Crystal structures and structural stability of Cs$_{0.5}$MA$_{0.5}$PbI$_3$, Cs$_{0.75}$MA$_{0.25}$PbI$_3$ and Cs$_{0.25}$MA$_{0.75}$PbI$_3$; TB parameters for Cs$_{0.5}$MA$_{0.5}$PbI$_3$. 
	\end{suppinfo}

\providecommand{\latin}[1]{#1}
\makeatletter
\providecommand{\doi}
{\begingroup\let\do\@makeother\dospecials
	\catcode`\{=1 \catcode`\}=2 \doi@aux}
\providecommand{\doi@aux}[1]{\endgroup\texttt{#1}}
\makeatother
\providecommand*\mcitethebibliography{\thebibliography}
\csname @ifundefined\endcsname{endmcitethebibliography}
{\let\endmcitethebibliography\endthebibliography}{}